\documentclass[twocolumn,nofootinbib,amsmath,prl,aps,superscriptaddress,tightenlines,preprintnumbers]{revtex4}

\usepackage{graphicx}
\usepackage{braket}
\usepackage[usenames]{color}
\usepackage{latexsym}
\usepackage{array,multirow}
\usepackage{amsmath}
\usepackage{amsfonts}
\usepackage{slashed}
\usepackage[normalem]{ulem}
\usepackage{calligra}
\usepackage{bm}
\usepackage{mathrsfs}
\usepackage[caption=false]{subfig}

\usepackage{tikz}
\usetikzlibrary{arrows,decorations.pathmorphing,backgrounds,positioning,fit,petri,automata,shadows,calendar,mindmap,
decorations.markings,calc}

\usepackage[utf8]{inputenc}
\usepackage{graphicx}
\usepackage{feynmp-auto}

\def\bea{\begin{eqnarray}}
\def\eea{\end{eqnarray}}
\def\hyp{\mathsf{y}}

\newcommand{\ov}{\overline}

\newcommand{\m}{{\ov m}}
\def\nn{\nonumber\\ }

\newcommand{\Lagr}{\mathcal{L}}

\begin{document}

\newcount\hour \newcount\minute
\hour=\time \divide \hour by 60
\minute=\time
\count99=\hour \multiply \count99 by -60 \advance \minute by \count99
\newcommand{\mydate}{\ \today \ - \number\hour :00}

\title{Equations of Motion for the Standard Model Effective Field Theory:\\ Theory and Applications.}

\author{Abdurrahman Barzinji, Michael Trott and Anagha Vasudevan,\\
Niels Bohr International Academy,
University of Copenhagen, \\
Blegdamsvej 17, DK-2100 Copenhagen, Denmark
}

\begin{abstract}
The equations of motion for the Standard Model Effective Field Theory (SMEFT) differ from those in the Standard Model.
Corrections due to local contact operators modify the equations of motion and impact matching results at sub-leading order
in the operator expansion. As a consequence, a matching coefficient in $\mathcal{L}^{(n)}$ (for operators of dimension $n$) can be
dependent on the basis choice for $\mathcal{L}^{(m<n)}$. We report the SMEFT equations of motion with corrections
due to $\mathcal{L}^{(5,6)}$. We demonstrate the effect of these corrections when matching to sub-leading order by
considering the interpretation of recently reported $B \to K^{(*)} \ell^+ \ell^-$ lepton
universality anomalies in the SMEFT.
\end{abstract}

\maketitle
\newpage

\paragraph{\bf I. Introduction.}

When physics beyond the Standard Model (SM) is present at scales
$\Lambda > \sqrt{2 \, \langle H^\dagger H\rangle} = \bar{v}_T$,
the SM can be extended into an effective field theory (EFT).
Such an EFT can be constructed
with two further defining assumptions:
no light hidden states in the spectrum with couplings
to the SM; and a $\rm SU(2)_L$ scalar doublet with hypercharge
$\hyp_h = 1/2$ being present in the EFT.
The resulting Standard Model Effective Field Theory (SMEFT) extends the SM with higher dimensional operators $\mathcal{Q}_i^{(\rm{d})}$
of mass dimension $\rm{d}$:
\begin{align}
	\Lagr_{\textrm{SMEFT}} &= \Lagr_{\textrm{SM}} + \Lagr^{(5)}+\Lagr^{(6)} +
	\Lagr^{(7)} + \dots,  \\ \nonumber \Lagr^{(\rm{d})} &= \sum_i \frac{C_i^{(\rm{d})}}{\Lambda^{\rm{d}-4}}\mathcal{Q}_i^{(\rm{d})}
	\textrm{ for } \rm{d}>4.
\end{align}
The operators $Q_i^{(\rm{d})}$ are suppressed by $\rm{d}-4$ powers of the cut-off scale $\Lambda$ and
the $C_i^{(\rm{d})}$ are the Wilson coefficients. We use the non-redundant $\mathcal{L}^{(6)}$ Warsaw basis \cite{Grzadkowski:2010es}, which
removed some redundancies (see also \cite{AguilarSaavedra:2010zi}) in
the overcomplete basis of Ref.~\cite{Buchmuller:1985jz}.

The exact size of the SMEFT expansion parameters: $\overline{v}_T^2/\Lambda^2 < 1$, $p^2/\Lambda^2 <1$ ($p^2$ stands for a general dimension two kinematic Lorentz invariant, $\overline{v}_T^2$ is the modified Higgs potential\footnote{In the SMEFT, the Higgs scalar doublet potential is modified by the inclusion of the operator $\mathcal{Q}_H=(H^{\dagger}H)^3$ yielding a new minimum $\braket{H^{\dagger} H}=\frac{v^2}{2}(1+\frac{3 C_H v^2}{4 \lambda})=\frac{1}{2} v_{T}^2$, see Ref. \cite{Alonso:2013hga}}),  are unknown and modified by the
$C_i^{(\rm{d})}$. As a result when deviations from the SM are interpreted in the SMEFT formalism, sub-leading results
and loop corrections are sometimes of interest in interpreting a experimental result.

To perform a matching to a non-redundant operator basis for $\mathcal{L}^{(\rm{d})}$,
it is usually required to know the Equations of Motion (EOM),
including possible SMEFT corrections due to $\mathcal{L}^{(n<\rm{d})}$.
In this paper, we report the EOM for the SMEFT including corrections due to
$\mathcal{L}^{(5,6)}$ and demonstrate the utility of these results in some
examples.

A partial discussion concerning corrections of this form has recently appeared in literature in Refs.~\cite{Elgaard-Clausen:2017xkq,Jenkins:2017dyc}.
Ref.~\cite{Elgaard-Clausen:2017xkq} discusses EOM corrections to matching results of the seesaw model to sub-leading order,
while Ref.~\cite{Jenkins:2017dyc} discusses the importance of these corrections to matching between the SMEFT and the low-energy EFT
where some Standard Model particles are integrated out.

\paragraph{\bf II. Notation and conventions.}\label{NotationDefinitions}
The SM Lagrangian \cite{Glashow:1961tr,Weinberg:1967tq,Salam:1968rm} notation is fixed to be
\bea\label{sm1}
\mathcal{L} _{\rm SM} &=& -\frac14 G_{\mu \nu}^A G^{A\mu \nu}-\frac14 W_{\mu \nu}^I W^{I \mu \nu} -\frac14 B_{\mu \nu} B^{\mu \nu}, \\
&\,&\hspace{-0.75cm}  + \sum_{\psi} \overline \psi\, i \slashed{D} \, \psi + (D_\mu H)^\dagger(D^\mu H) -\lambda \left(H^\dagger H -\frac12 v^2\right)^2\nonumber \\
&-& \biggl[ H^{\dagger j} \overline d\, Y_d\, q_{j}
+ \widetilde H^{\dagger j} \overline u\, Y_u\, q_{j} + H^{\dagger j} \overline e\, Y_e\,  \ell_{j} + \hbox{h.c.}\biggr], \nonumber
\eea
where $\psi = \{q,\ell,u,d,e \}$ are four component Dirac spinors that transform as $\{\bf 2,2,1,1,1 \}$ under $\rm SU(2)_L$.
The fermion fields $q$ and $\ell$ are left-handed fields and transform as $(1/2,0)$ under
the restricted Lorentz group $\rm SO^+(3,1)$. The $u$, $d$ and $e$ are right-handed fields and transform as
$(0,1/2)$. The chiral projectors have the convention $\psi_{L/R} = P_{L/R} \, \psi$ where
$P_{R/L} = \left(1 \pm \gamma_5 \right)/2$.
The gauge covariant derivative is defined with a positive sign convention $
D_\mu = \partial_\mu + i g_3 T^A A^A_\mu + i g_2  t^I W^I_\mu + i g_1 {\bf \hyp_i} B_\mu$.
${\bf \hyp_i}$ is the $\rm U_Y(1)$ hypercharge generator. The $\rm SU_c(3)$ generators
($T^A$) are defined with normalization ${\rm Tr}(T^A T^B) = 2 \delta^{AB}$ and finally
$t^I=\tau^I/2$ are the $\rm SU_L(2)$ generators, with $\tau^I$ the Pauli matrices.
$\widetilde H_j = \epsilon_{jk} H^{\star\, k}$
where the $\rm SU_L(2)$ invariant tensor $\epsilon_{jk}$ is defined by $\epsilon_{12}=1$ and $\epsilon_{jk}=-\epsilon_{kj}$.
At times we raise or lower this index in notation for clarity on index sums.
The flavour indices are suppressed in Eqn.~(\ref{sm1}),
the fermion mass matrices are defined as $M_{u,d,e}=Y_{u,d,e}\, v /\sqrt 2$ with
$Y_{u,d,e},M_{u,d,e}$ each $ \subset \mathbb{C}_{3 \times 3}$ in flavour space.
Our conventions are consistent with the SMEFT review \cite{Brivio:2017vri} and further notational conventions are defined in the Appendix.

The leading correction to the SM violates Lepton number due to its operator dimension \cite{deGouvea:2014lva,Kobach:2016ami}.
We use for $\mathcal{L}^{(5)}$  the non-Hermitian operator \cite{Weinberg:1979sa,Wilczek:1979hc}
\bea\label{weinberg}
\mathcal{Q}_{mn}^{(5)} = \left(\overline{\ell^{c, m}} \, \tilde{H}^\star\right) \left(\tilde{H}^\dagger \, \ell^{n}\right),
\eea
with spinor conventions defined as follows.
The $c$ superscript corresponds to a charge conjugated Dirac four-component spinor $\psi^c  = C \overline{\psi}^T$
with $C= - i \gamma_2 \, \gamma_0$ in the chiral basis for the $\gamma_i$ we use. The star superscript is reserved for
complex conjugation operation that is applied to bosonic quantities. As chiral projection and $c$ do not commute we fix notation that
$\ell^c$ denotes a doublet lepton field chirally projected and subsequently charge conjugated.

\paragraph{\bf III. Formalism for EOM to sub-leading order.}\label{formalism}
The SMEFT Lagrangian is composed of a series of $d$ dimensional operators in $\Lagr^{(d)}$.
Counting the independent operators in a non-redundant operator basis for  $\Lagr^{(d)}$ can be
performed efficiently using the results of Refs.~\cite{Lehman:2015via,Lehman:2015coa,Henning:2015alf,Henning:2017fpj}.

Reducing a basis to a non-redundant form
using the EOM is related to the possibility to perform gauge independent field redefinitions that satisfy the equivalence
theorem of $S$-matrix elements
\cite{Chisholm:1961tha,Kamefuchi:1961sb,Coleman:1969sm,Kallosh:1972ap}.
The full set of all possible $\rm SU(3)_c \times SU(2)_L \times U(1)_Y$ preserving small field redefinitions
on the SM fields, collectively denoted $\mathcal{F}$, up to order $\rm{d}$, can be denoted
\bea\label{smallfield}
\mathcal{F} \rightarrow \mathcal{F}' + \sum_{i=1}^{\rm{d}} c_i \, \left(\frac{\mathcal{O}_i}{\Lambda^i}\right),
\eea
with $\mathcal{O},\mathcal{F}'$ both transforming as $\mathcal{F}$ under $\rm SU(3)_c \times SU(2)_L \times U(1)_Y$ and
subject to ${\rm dim} [\mathcal{F}] = {\rm dim} [\mathcal{O}_i] - {\rm dim}[\Lambda^i]$.
Redefining the field variables with a specific sequence of the full set of $\mathcal{F}$ transformations, a non-redundant basis can be defined by choosing $c_i$ to cancel the largest set of
operators possible in an overcomplete basis.\footnote{At $\mathcal{L}^{(5)}$ only $\mathcal{Q}_{mn}^{(5)}$ (and its Hermitian conjugate)
are present. It is interesting to note that, equivalently, for ${\rm dim}[\Lambda^i] = 1$,
no factorization of $\mathcal{O}_i$ is possible into a field variable
transforming as $\mathcal{F}$ while $\mathcal{O}_i/\mathcal{F}$ is composed of
dynamical (SMEFT symmetry preserving) fields.}. For a more thorough discussion on field redefinitions and the removal of redundant operators, see Ref.~\cite{Brivio:2017vri}.

Consistency conditions result from this procedure.
Some of these conditions are the EOM relations between operators of different bases.
Another consequence is that the higher dimensional operators play a role in the renormalization group evolution of the Lagrangian parameters of dimension $d \leq 4$. For the Warsaw
basis, the RG running results of this form were reported in Ref.~\cite{Jenkins:2013zja}.

In this paper we address another set of consistency conditions, the modifications of the EOM in a particular operator basis.
Once $\mathcal{L}_{SMEFT}$ is defined up to dimension
${\rm (n)}$, when considering matching up to this canonical dimension, the higher dimensional operators
themselves correct the SM EOM due to operators of dimension $m < n$. This results in matching results at dimension ${\rm (n)}$
having a subtle dependence on basis choice at order $m<n$. This effect comes about as the field variables themsleves
are redefined in the EFT when defining a non-redundant operator basis.

The EOM for the SMEFT, as in the SM, are defined by the condition that the variation of the action with respect to the
fields vanishes ($\delta S = 0$), where
\bea
S = \int \mathcal{L}_{SMEFT}(\mathcal{F}, D_\mu \mathcal{F}) \, d^4 x,
\eea
resulting in
\bea\label{eulerlagrange}
0 = \int d^4 x \left[\frac{\partial \mathcal{L}_{SMEFT}}{\partial \mathcal{F}} \delta \mathcal{F} - \partial_\mu (\frac{\partial  \mathcal{L}_{SMEFT}}{\partial  (\partial_\mu \mathcal{F})}) \,  \delta \mathcal{F}\right],
\eea
where the surface term given by
\bea
\partial_\mu \left(\frac{\partial  \mathcal{L}_{SMEFT}}{\partial (\partial_\mu \mathcal{F})} \, \delta \mathcal{F} \right),
\eea
vanishes. The surface term vanishes up to an accuracy dictated
by the power counting of the SMEFT to order ${\rm (d)}$, as this is the accuracy to which
the field variables are defined.
The surface term and variation are defined by a partial derivative.
At low orders in the operator dimension expansion of $\mathcal{L}_{SMEFT}$ the EOM terms are simplified into a form with
covariant derivatives in the adjoint and fundamental representations due to renormalizability.
This simplification is present in the SM EOM, but is not present in the SMEFT EOM corrections in some cases, as shown below.

\paragraph{\bf IV. SMEFT EOM.}\label{examples}
We use the Hermitian derivative conventions and integration by parts identity
\bea
H^\dagger \, i\overleftrightarrow D_\mu H &=& i H^\dagger (D_\mu H) - i (D_\mu H)^\dagger H, \\
H^\dagger \, i\overleftrightarrow D_\mu^I H &=& i H^\dagger \tau^I (D_\mu H) - i (D_\mu H)^\dagger \tau^I H, \\
\mathcal{Q}_{H \Box} + 4 \, \mathcal{Q}_{HD} &=&   (H^\dagger \, i\overleftrightarrow D^\mu H)(H^\dagger \, i\overleftrightarrow D_\mu H).
\eea
The currents of the SM fields are defined as
\begin{align}
j_\mu^A &=\sum_{\psi=u,d,q} \overline \psi \, T^A \gamma_\mu  \psi\,, \\
j_\mu^I &= \frac 12 \overline q \, \tau^I \gamma_\mu  q + \frac12 \overline \ell \, \tau^I \gamma_\mu  \ell + \frac12 H^\dagger \, i\overleftrightarrow D_\mu^I H, \\
j_\mu &=\sum_{\psi=u,d,q,e,\ell} \overline \psi \, \hyp_i \gamma_\mu  \psi + \frac12 H^\dagger \, i\overleftrightarrow D_\mu H.
\end{align}
Corrections to the SM EOM gauge fields are
\begin{align}
\left[D^\mu , G_{\mu \nu} \right]^A &= g_3 \, j_\nu^A + g_3 \sum_{\rm{d}=5}^{\rm{\infty}} \frac{\Delta^{A,(\rm{d})}_{G,\nu}}{\Lambda^{\rm{d-4}}}, \nn
\left[D^\mu , W_{\mu \nu} \right]^I &= g_2  j_\nu^I + g_2 \sum_{\rm{d}=5}^{\rm{\infty}} \frac{\Delta^{I,(\rm{d})}_{W,\nu}}{\Lambda^{\rm{d-4}}}, \nn
D^\mu B_{\mu \nu} &= g_1  j_\nu + g_1 \sum_{\rm{d}=5}^{\rm{\infty}} \frac{\Delta_{B,\nu}^{(\rm{d})}}{\Lambda^{\rm{d-4}}}.
\end{align}
$\Delta^{(6)}$ here contains the full set of corrections to each field's EOM, due to the complete Warsaw basis of $\mathcal{L}^{(6)}$ operators.
These corrections are reported in the Appendix.

The covariant derivatives for an operator $\mathcal{Q}$ in the adjoint representations of $\rm SU(2)$ and $\rm SU(3)$ are
\bea
\left[D^\mu , \mathcal{Q} \right]^I = \partial^\mu \, \mathcal{Q}^I - g_2 \, \epsilon^{JKI} \, W^{\mu}_J \mathcal{Q}_K, \\
\left[D^\mu , \mathcal{Q} \right]^A = \partial^\mu \, \mathcal{Q}^A - g_3 \, f^{BCA} \, A^{\mu}_B \mathcal{Q}_C.
\eea
Corrections to the SM EOM for the fermions are of the form (colour indices are suppressed)
\begin{align}
i\slashed{D}\, q^j_m &= u^n \, [Y_u]_{n m }^\star\, \widetilde H^j + d^n \,[Y_d]_{n m}^\star\, H^j  +
\sum_{\rm{d}=5}^{\rm{\infty}} \frac{\Delta^{j,(\rm{d})}_{q, m}}{\Lambda^{\rm{d-4}}}, \nn
i\slashed{D} \, \ell_m^j &= [Y_e]_{n m}^\star \, e^n  H^j + \sum_{\rm{d}=5}^{\rm{\infty}} \frac{\Delta_{\ell, m}^{j,({\rm{d}})}}{\Lambda^{\rm{d-4}}}, \nn
i\slashed{D}\,  d_m &= [Y_d]_{m n}\, q_{n}^j \, H^{\dagger}_j + \sum_{\rm{d}=5}^{\rm{\infty}} \frac{\Delta^{(\rm{d})}_{d, m}}{\Lambda^{\rm{d-4}}},\nn
i\slashed{D}\, u_m &= [Y_u]_{m n}\, q^n_j\, \widetilde H^{\dagger\, j} + \sum_{\rm{d}=5}^{\rm{\infty}} \frac{\Delta_{u, m}^{(\rm{d})}}{\Lambda^{\rm{d-4}}},\nn
i\slashed{D} \, e_m &= [Y_e]_{m n}\, \ell^n_j H^{\dagger\, j} + \sum_{\rm{d}=5}^{\rm{\infty}} \frac{\Delta_{e, \kappa}^{(\rm{d})}}{\Lambda^{\rm{d-4}}}.
\end{align}
The modifications of the Higgs EOM in the SMEFT are
\bea
D^2 H^j &=& \lambda v^2 H^j  -2 \lambda (H^\dagger H) H^j
 - \overline q_k^n \, [Y_u]_{m n}^\star\, u^m \epsilon^{kj}, \nn
&-& \overline d^n  [Y_d]_{n m} q_m^j
- \overline e^n  [Y_e]_{n m}  \ell^{m, j} + \sum_{\rm{d}=5}^{\rm{\infty}} \frac{\Delta^{j,({\rm d})}_H}{\Lambda^{\rm{d-4}}}.
\eea
The corrections for $\mathcal{L}^{(5)}$ using Eqn.~\ref{weinberg} are
\bea
\Delta_{\ell,m}^{j,(5)} &=& - 2 \, C_{n m}^{(5) \, \star} \, \tilde{H}^j \,  \left( \tilde{H}^T \ell^{c}_{n}\right), \\
\Delta_H^{j,(5)} &=&  - C_{nm}^{(5) \, \star} \, \epsilon^{jk} \,
\left[\overline{\ell^{m}_k} \, (\tilde{H}^T \, \ell^c_n) + (\overline{\ell^m} \tilde{H}) \, \ell_n^{c,k} \right]
\eea

\paragraph{\bf V. Matching examples}\label{examples}
As an illustrative set of examples of matching using the SMEFT EOM, we consider the interpretation of anomalous measurements of $B \to K^{(*)} \ell^+ \ell^- $
lepton universality ratios for $\ell_{m=\{1,2\}} = \{e, \mu \}$ \cite{Aaij:2014ora,Aaij:2017vbb} which have shown some minor tension with the SM predictions.
Such anomalies could signal physics beyond the SM\footnote{These anomalies could also be statistical fluctuations, as indicated by their global (in)significance.
Here our interest in these anomalies only extends to an illustrative example of EOM SMEFT effects.} that induce the $\mathcal{L}^{(6)}$
operators
\bea\label{anomalyoperators}
\mathcal{Q}^{(1)}_{\substack{lq\\ mm sb}} &=& (\bar{\ell}_m \gamma^\mu \ell_m) (\bar{s} \gamma_\mu b),\\
\mathcal{Q}^{(3)}_{\substack{lq\\ mm sb}} &=& (\bar{\ell}_m \, \tau^I \, \gamma^\mu \ell_m) (\bar{s} \, \tau_I \, \gamma_\mu b).
\label{anomalyoperators2}
\eea
The operators and anomalies of interest can come about
by matching at tree level to $\mathcal{L}^{(6)}$ the effect of fields denoted as
$\{\zeta,\beta,\mathcal{W}, \mathcal{U}_2, \chi \}$ (using the notation of Ref.~\cite{deBlas:2017xtg}), for example.
These fields have the $\{\rm{SU}(3),\rm{SU}(2) \}_{U_Y(1)}$ representations, with the spin of each field given as a superscript
\bea
\{(3,3)^{0}_{-1/3}, \, (1,1)^{1}_0, \,(1,3)^{1}_0, \, (3,1)^{1/2}_{2/3}, \, (3,3)^{1/2}_{2/3}\}.
\eea

The field $\zeta$ leads to the baryon number violating operator
$\mathcal{Q}_{qqq}$, indicating a very small matching coefficient. In addition, the
operators $\mathcal{Q}_{qq}^{(1,3)},\mathcal{Q}_{\ell q}^{(1,3)}$ are also induced in a tree level matching. Since $\zeta$ is a scalar field,
the low-momentum expansion of a scalar propagator introduces a dependence on the momentum $p$ flowing through the scalar propagator at subleading order- i.e.,
$p^2/m_\zeta^4$, which can be reduced using EOM. The first irreducible corrections appear only at $\mathcal{L}^{(10)}$.

\begin{figure*}[t]
\includegraphics[scale=0.25]{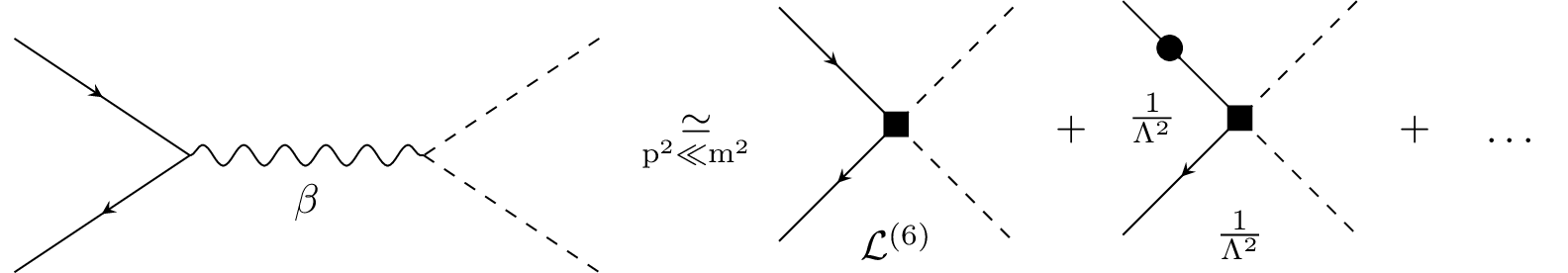}
\caption{The black box indicates the small momentum expansion of the propagator $\beta$ in terms of local operators. The EOM matching correction of a singlet field $(1,1)_0$
is then indicated as a filled circle on the field.} \label{fig1}
\end{figure*}

The heavy vector fields $\{\beta,\mathcal{W} \}$ are more interesting when considering EOM corrections in $\mathcal{L}_{SMEFT}^{(8)}$.
Consider the singlet field $\beta$, with a bare mass introduced via the Stueckelberg mechanism \cite{Stueckelberg:1938zz},
as encoded in a Proca Lagrangian. The $\beta$ field is coupled to the SM through
\bea
&\mathcal{L}_{int}^\beta =  - g^{R H}_\beta \, \beta^\mu H^\dagger i \overleftrightarrow D_\mu H
+ g^{I H}_\beta \beta^\mu \partial_\mu (H^\dagger H), \nn
&- \hspace{-0.5cm} \sum\limits_{\psi = \{\ell,q,e,d,u \}} \hspace{-0.5cm} g_{\beta \, \psi}^{mn} \, \beta^\mu \, \bar{\psi}_m \gamma_\mu \psi_n.
\eea
Here $g^{R H}_\beta$ and $g^{I H}_\beta$ are real and  imaginary components of  the coupling of the $\beta$
field to the non-Hermitian scalar current $H^\dagger i D^\mu H$.
Integrating out $\beta$ gives the $\mathcal{L}^{(6)}$ Wilson coefficients
\begin{align}
C_{\substack{H \psi_1 \\ mp}}^{(1)} &= -g^{R H}_\beta \, g^{\psi_1}_{mp}, &  {C_{\substack{\psi_1 \, \psi_2 \\ mpst}}} &= -g_{\beta \psi_1}^{mp} \, g^{\psi_2}_{st},
\end{align}
directly, here $\psi_1 \neq \psi_2$, in the case of $\psi_1 = \psi_2$, a further factor of two is present in $C_{\psi_1 \, \psi_2}$.
When using the Warsaw basis to define the matching to $\mathcal{L}^{(6)}$, products of currents are reduced with the EOM
and integration by parts. The latter is used to simplify the pure Higgs currents into
\begin{align}
C_{H \Box} &= - \frac{(g^{R H}_\beta)^2}{2} + \frac{(g^{I H}_\beta)^2}{2}
 & C_{HD} &= - 2\, (g^{R H}_\beta)^2.
\end{align}

These matching results have been verified against the comprehensive tree-level matching dictionary given in Ref.~\cite{deBlas:2017xtg}.

In addition, the following products of currents are also reduced with the EOM
\bea
\mathcal{L}^{(6)}_\beta  &\supset& \frac{1}{m_\beta^2} \, g^{R H}_\beta \, g^{I H}_\beta (H^\dagger i \overleftrightarrow D_\mu H) \, \partial^\mu (H^\dagger H), \nn
&+&  \frac{g^{I H}_\beta}{m_\beta^2} \, \partial_\mu (H^\dagger H) \hspace{-0.5cm} \sum\limits_{\psi = \{\ell,q,e,d,u \}} \hspace{-0.5cm} g_{\beta \, \psi}^{mn} \, \bar{\psi}_m \gamma^\mu \psi_n.
\eea
This generates the Wilson coefficients
\bea
C_{\substack{e H\\ pr}} &= -i g^{I H}_\beta\left([Y_e]_{rp}^\star \, g^{R H}_\beta
- [Y_e]_{rm}^\star \,  g_{\beta, \ell}^{pm} + [Y_e]_{mp}^\star \, g_{\beta, e}^{mr}\right), \nn
C_{\substack{d H\\ pr}} &= -i g^{I H}_\beta\left([Y_d]_{rp}^\star \, g^{R H}_\beta
- [Y_d]_{rm}^\star \,  g_{\beta, q}^{pm} + [Y_d]_{mp}^\star \, g_{\beta, d}^{mr}\right), \nn
C_{\substack{u H\\ pr}} &= -i g^{I H}_\beta\left( [Y_u]_{rp}^\star \, g^{R H}_\beta
- [Y_u]_{rm}^\star \,  g_{\beta q}^{pm} + [Y_u]_{mp}^\star \, g_{\beta u}^{mr}\right), \nn
\eea
and their Hermitian conjugates. EOM corrections due to $\mathcal{L}^{(6)}$ are introduced into the matching
due to this procedure. In general a very large number of $\mathcal{L}^{(8)}$ matching corrections are introduced
in the SMEFT, as can be directly verified.

These matching contributions are non-intuitive (for the authors). They correspond to
the effect of redefining the field variables to fix the operator basis at $\mathcal{O}(1/\Lambda^2)$
in conjunction to tree level matching, as illustrated in Fig.~\ref{fig1}.
It is interesting to note that for this reason, the standard naive example of expanding a massive vector propagator
in $p^2/m^2$ to obtain a series of local contact operators to introduce the idea of EFT, is quite an incomplete
description of the physics defining the SMEFT at sub-leading order.\footnote{Of course, other effects at sub-leading order also exist, including
the expansion of the matrix elements in the power counting. This point, consistent with the content of this work, and Refs.~\cite{Elgaard-Clausen:2017xkq,Jenkins:2017dyc} was recently
stressed by Ref.~\cite{Criado:2018pv}. In addition, Ref.~\cite{Criado:2018pv} [v1] mischaracterizes the content and claims
of this work and Ref.\cite{Jenkins:2017dyc}, as any reader can directly confirm.}

\begin{figure}
	\includegraphics[scale=0.25]{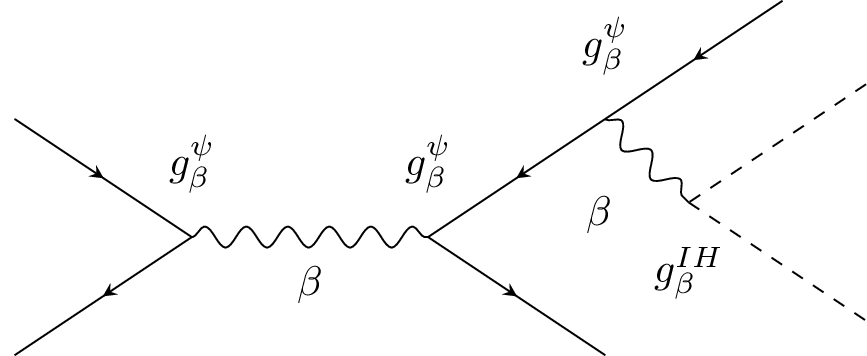}
\caption{Feynman diagram with the scaling of couplings obtained in the EOM correction.
Note the necessary presence of a light internal intermediate propagating state.} \label{fig2}
		\end{figure}

Restricting our attention to the corrections due to the operator
in Eqn.~(\ref{anomalyoperators}) one finds the matching corrections
\bea\label{results}
\mathcal{L}^{(8)}_\beta &\supset& \frac{i g_\beta^{IH}  g_{\beta \ell}^{mn}}{m_\beta^4}
\left[C_{\substack{\ell q \\ n pst}}^{(1)} J^{\ell}_{mp, \mu}
 -
 C_{\substack{\ell q \\ p m st}}^{(1)} J^{\ell}_{pn, \mu} \right] J^{q, \mu}_{st} H^\dagger H, \nn
&+& \frac{i g_\beta^{IH}  g_{\beta \,q}^{mn}}{m_\beta^4}
\left[C_{\substack{\ell q \\ stnp}}^{(1)} J^{q}_{mp, \mu} -
 C_{\substack{\ell q \\ stpm}}^{(1)}
 J^{q}_{pn, \mu} \right] J^{\ell, \mu}_{st} H^\dagger H. \nn
\eea
Definition of $J^{\psi}_{pr, \, \mu}$ is given by \ref{eq:Jdef} in the Appendix. The scaling of these matching contributions with couplings to $\beta$ are also non-intuitive.
A directly constructed Feynman diagram
with this coupling scaling involves two intermediate $\beta$ fields and an internal propagator
of the light states retained in the SMEFT, as shown in Fig.~\ref{fig2}. The light
intermediate state propagator leads to the lack of a local operator in the low momentum limit
defining the SMEFT. Corrections with this coupling dependence
are nevertheless still present as local contact operators in the SMEFT, they come about due to the
contributions illustrated
in Fig.~\ref{fig1}.

The presence of an explicit factor of $i$ in Eqn.~(\ref{results}) indicates that
the decomposition of the Wilson coefficient of the operator into real and imaginary components leads to the cancellation of
some terms symmetric in the flavour indices. Recall that the flavour indices that are bi-linear
in the same field of a self Hermitian operator can be decomposed into a real symmetric $S_{pr}$ (CP-even) and real anti-symmetric $A_{pr}$
(CP-odd) dependence as
\bea
C_{pr} = S_{pr} + i A_{pr}.
\eea
The anti-symmetric components of the Hermitian operator's Wilson coefficient do not cancel in the case considered,
but the symmetric components do cancel. Such cancellations occur as the derivative terms reduced
with the EOM in defining the Warsaw basis act on bi-linear currents with the same fermion field.
As the $B$ meson anomalies are associated with flavour
diagonal lepton interactions, but off-diagonal quark flavour indices, the second term in Eqn.~\ref{results}
survives for case of the $\beta$ field leading to these anomalies with a CP-odd phase.
This effect comes about directly when the $\beta$ field is promoted to a complex singlet vector field.

The EOM effects at sub-leading order due to the heavy field $\mathcal{W}$ are similar.
The interaction Lagrangian with the SM fields is then
\bea\label{theeffect}
& 2 \, \mathcal{L}_{int}^\mathcal{W} =  - g^{R H}_\mathcal{W} \, \mathcal{W}_I^\mu H^\dagger i \overleftrightarrow D^I_\mu H
+ g^{I H}_\mathcal{W} \mathcal{W}_I^\mu \partial_\mu (H^\dagger \tau^I H), \nn
&- \hspace{-0.25cm} \sum\limits_{\psi = \{\ell,q\}} \hspace{-0.2cm} g_{\mathcal{W} \psi}^{mn} \, \mathcal{W}_I^\mu \, \bar{\psi}_m \tau^I \, \gamma_\mu  \psi_n,
\eea
leading to the contact operators
\bea
\mathcal{L}^{(6)}_\mathcal{W} &\supset& \frac{1}{4 \, m_\mathcal{W}^2} \, g^{R H}_\mathcal{W} \, g^{I H}_\mathcal{W} (H^\dagger i \overleftrightarrow D^I_\mu H) \, \partial^\mu (H^\dagger \tau^I H), \nn
&+&  \frac{g^{I H}_\mathcal{W}}{4 \, m_\mathcal{W}^2} \, \partial^\mu (H^\dagger \tau^I H) \hspace{-0.25cm} \sum\limits_{\psi = \{\ell,q \}} \hspace{-0.25cm} g_{\mathcal{W} \psi}^{mn} \, \bar{\psi}_m \tau^I \gamma_\mu \psi_n,
\eea

This results in corrections to $\mathcal{L}_{\mathcal{W}}^{(8)}$ due to the field $\mathcal{W}$, which include the operator in Eqns.~(\ref{anomalyoperators2}).
The matching is analogous to the form in Eqn.~\eqref{results} with the operators $Q_{\ell q}^{(3)}$ replacing the operators $Q_{\ell q}^{(1)}$. In addition, the normalization differs by a factor of 4.

The fermion fields $\{ \mathcal{U}_2, \chi\}$  do not lead to an EOM reduction in $\mathcal{L}^{(6)}$
when using the Warsaw basis. As such, they do not induce non-intuitive corrections through the EOM of this form.
This result is due to the particular representations that these fermion fields carry and is not general.
For example, integrating out a singlet fermion field in Ref.~\cite{Elgaard-Clausen:2017xkq}
leads to sub-leading corrections at $\mathcal{L}^{(7)}$ due to EOM reductions of $\mathcal{L}^{(6)}$
when considering the matching of the minimal seesaw model to $\mathcal{L}_{SMEFT}$.

\paragraph{\bf VI. Conclusions}\label{conclusions}
In this paper we have determined the corrections due to $\mathcal{L}^{(5)}$
and $\mathcal{L}^{(6)}$ to the SMEFT equations of motion. These corrections introduce
a dependence on the operator basis defined at dimension $\mathcal{L}^{(6)}$ when matching
to $\mathcal{L}^{(7)}$, or higher orders. Incorporating EOM corrections can be essential to correctly determine
matching to sub-leading order in the SMEFT. We have illustrated
these effects in some simple matching examples using the $B$ meson anomalies as motivation.
\paragraph{\bf Acknowledgments}

\begin{acknowledgments}
The authors acknowledge support from the Villum Fonden and the Danish National Research Foundation (DNRF91)
  through the Discovery center. MT thanks M.I.T.P. and the organizers of BSMPR2018 for support
  while part of this work was completed. We thank Ilaria Brivio, Andreas Helset
	and E.Bjerrum-Bohr for useful discussions and comments on the manuscript.
\end{acknowledgments}
\paragraph{\bf Bibliography}\label{bib}
\bibliography{bibliography_V3}
\newpage
\begin{widetext}
\appendix*  \begin{center}{\bf Appendix} \end{center}
We use the notation
\begin{align}\label{eq:Jdef}
J^{\psi \, \mu}_{pr} &= \overline{\psi}_p \gamma^\mu \psi_r, \quad &
J^{\psi, I, \, \mu}_{pr} &= \overline{\psi}_p \gamma^\mu \, \tau^I \, \psi_r, \quad &
J^{\psi, A, \, \mu}_{pr} &= \overline{\psi}_p \gamma^\mu \, T^A \, \psi_r, \\
\mathcal{C}^{\mu \nu}_{\substack{\psi_1  \psi_2 F \\pr, {\bf{g}}}} &= C_{\substack{\psi_1 F\\pr}} \, \overline{\psi_{2,p}} \sigma_{\mu \nu} {\bf{G}} \, \psi_{1,r} H + {\rm h.c.}, \quad &
\tilde{\mathcal{C}}^{\mu \nu}_{\substack{\psi_1  \psi_2 F \\pr, {\bf{g}}}} &= C_{\substack{\psi_1 F\\pr}} \, \overline{\psi_{2,p}} \sigma_{\mu \nu} {\bf{G}} \psi_{1,r} \tilde{H} + {\rm h.c.},
\end{align}
with $({\bf{g}},{\bf{G}}) = \{( \, ,\mathbb{I}),(I,\tau^I),(A,T^A)\}$ for more compact expressions. \\
Throughout the paper, we use the following convention for the subscripts to denote quantum numbers. $\rm{SU}(3)_c$ indices are $\{\alpha, \beta ,\rho\}$ and the corresponding generator indices are $\{A,B,C,D,E\}$. The $\rm{SU}(2)$ indices are represented by $\{i,j,k,l,o\}$ and its generator indices by $\{I,J,K,R,S\}$. The Lorentz indices are represented by the Greek letters $\{\mu, \nu ,\gamma, \phi\}$ and lastly the flavor (fermion family) indices are denoted by $\{p, r, s, t, m, n\}$ .

The corrections due to $\mathcal{L}^{(6)}$ in the "Warsaw" basis for the Higgs EOM are

\begin{align}
\Delta_{H}^{j,(6)} &= 3  C_H (H^\dagger H)^2 H^j + \hspace{-0.5cm}
\sum_{F =\{W,B,G\}} \hspace{-0.5cm} \left(C_{HF} \, H^j F_{\mu \nu} F^{\mu \nu} + C_{H\tilde{F}} \, H^j \tilde{F}_{\mu \nu} F^{\mu \nu}\right)
+ \hspace{-0.35cm} \sum_{F =\{W,\tilde{W}\}} \hspace{-0.5cm} C_{HFB} \,(\tau_I H)^j F^I_{\mu \nu} B^{\mu \nu}, \nn
&+ \left(C_{\substack{uG \\pr}} \, \overline{q}_{p,k} \sigma_{\mu  \nu} T^A u_r \epsilon^{kj}
+ C_{\substack{dG \\pr}}^\star \, \overline{d}_r \sigma_{\mu  \nu} T^A q_p^j \right) G_A^{\mu \nu}
+ \left(C_{\substack{eB \\pr}}^\star \,\overline{e}_r \sigma^{\mu \nu} \ell^j_p
+ C_{\substack{uB \\pr}} \, \overline{q}_{p,k} \sigma_{\mu  \nu} u_r \epsilon^{kj}
+ C_{\substack{dB \\pr}}^\star \, \overline{d}_r \sigma_{\mu  \nu} \, q_p^j \right) B_{\mu \nu}, \nn
&+ \left(C_{\substack{eW \\pr}}^\star \overline{e}_r \sigma^{\mu  \nu}  \tau_I \ell^j_p
+ C_{\substack{uW \\pr}} \, \overline{q}_{p,k} \sigma_{\mu  \nu} \tau^I u_r \epsilon^{kj}
+ C_{\substack{dW \\pr}}^\star \, \overline{d}_r \sigma_{\mu  \nu} \tau_I q_p^j \right) W^I_{\mu \nu}
+ C_{\substack{eH \\pr}} \, H^j \,  \overline{\ell}_p e_r H
+ C_{\substack{eH \\pr}}^\star \, H^j \, (\overline{e}_r H^\dagger \ell_p), \nn
&+ C_{\substack{eH \\pr}}^\star \, (H^\dagger H) \,  \overline{e}_r \ell_p^j
+ C_{\substack{dH \\pr}} \, H^j \,  \left(\overline{q}_p d_r H\right)
+ C_{\substack{dH \\pr}}^\star \, H^j \, (H^\dagger \overline{d}_r q_p)
+ C_{\substack{dH \\pr}}^\star \, (H^\dagger H) \,  \overline{d}_r q_p^j
+ C_{\substack{uH \\pr}} \, H^j \,  \left(\overline{q}_p u_r \tilde{H}\right), \nn
& + C_{\substack{uH \\pr}}^\star\, H^j \, (\tilde{H}^\dagger \overline{u}_r q_p) + C_{\substack{uH \\pr}} (H^\dagger H)  \overline{q}_{p,k} u_r \epsilon^{kj}
+ 2 i (D_\mu H^j) \hspace{-0.2cm} \sum_{\psi = \{e,u,d\}} C_{\substack{H \psi \\ pr}} \, J^{\psi \, \mu}_{pr}
+  2 i (D_\mu H^j) \sum_{\psi = \{\ell,q\}} \, C_{\substack{H \psi \\ pr}}^{(1)} \, J^{\psi \, \mu}_{pr}, \nn
&+ C_{\substack{H \ell \\ pr}}^{(1)} \, H^j [Y_e]^\star_{mr} (\overline{\ell}_p e^m H)- C_{\substack{H \ell \\ pr}}^{(1)} \, H^j [Y_e]_{sp} (H^\dagger \overline{e}_s \ell^r)
+C_{\substack{H q \\ pr}}^{(1)} \, H^j [Y_u]^\star_{mr} (\overline{q}_p u^m \tilde{H})
- C_{\substack{H q \\ pr}}^{(1)} \, H^j [Y_u]_{mp} (\overline{u}_m q^r \tilde{H}^\dagger), \nn
&+C_{\substack{H q \\ pr}}^{(1)} \, H^j [Y_d]^\star_{mr} (\overline{q}_p d^m H)- C_{\substack{H q \\ pr}}^{(1)} \, H^j [Y_d]_{mp} (H^\dagger \, \overline{d}_m q^r)
+ C_{\substack{H e \\ pr}} \, H^j [Y_e]_{rm} (H^\dagger \overline{e}_p \ell^m)
- C_{\substack{H e \\ pr}} \, H^j [Y_e]^\star_{pm}(\overline{\ell}^m H e^r), \nn
&+ C_{\substack{H u \\ pr}} \, H^j [Y_u]_{rm} (\tilde{H}^\dagger \overline{u}_p \, q^m)
- C_{\substack{H u \\ pr}} \, H^j [Y_u]^\star_{pm}(\overline{q}^m \tilde{H} u^r)
+ C_{\substack{H d \\ pr}} \, H^j [Y_d]_{rm} (H^\dagger \overline{d}_p q^m)
- C_{\substack{H d \\ pr}} \, H^j [Y_d]^\star_{pm}(\overline{q}^m H d^r), \nn
&
+ i \, C_{\substack{H \ell \\ pr}}^{(3)}\left[\{\tau_I, D_\mu \} H \right]^j J^{\ell, I, \, \mu}_{pr} +i \, C_{\substack{H \ell \\ pr}}^{(3)} (\tau_I H)^j \partial_\mu \, J^{\ell, I, \, \mu}_{pr}
+ i \, C_{\substack{H q \\ pr}}^{(3)}
\left[\{\tau_I, D^\mu \} H \right]^j J^{q, I, \, \mu}_{pr} +
i \, C_{\substack{H q \\ pr}}^{(3)} (\tau_I H)^j \partial_\mu \, J^{q, I, \, \mu}_{pr}, \nn
&+ i \, C_{\substack{H ud \\ pr}}^\star \, (D_\mu \tilde{H})^j (\overline{d}_r \gamma^\mu u^p)
+ i \, C_{\substack{H ud \\ pr}}^\star \, \tilde{H}^j \partial_\mu \, (\overline{d}_r \gamma^\mu u^p)
- i \, C_{\substack{H ud \\ pr}}^\star \, (D_\mu H)^\dagger_k \epsilon^{kj} (\overline{d}_r \gamma^\mu u^p)
+ 2 \, C_{H \Box} \, H^j \Box (H^\dagger H), \nonumber \\
&- C_{HD}  \left[(D^\mu H)^j \,  (H^\dagger \overleftrightarrow D_\mu H) + H^j \partial^\mu  (H^\dagger \overrightarrow D_\mu H) \right].
\end{align}

The corrections to the fermion EOM due to the B number conserving $\mathcal{L}^{(6)}$ operators are
\begin{align}
\Delta_{e,m}^{(6,B)}
&= - C_{\substack{eH \\p m}}^\star \, (H^\dagger H) H^\dagger \, \ell_p
- C_{\substack{He \\m p}} \, (H^\dagger i\overleftrightarrow D_\mu H)\gamma^\mu e_p
- C_{\substack{eW \\ p m}}^\star \, \sigma^{\mu \nu} H^\dagger \, \tau_I \ell_p \, W^I_{\mu \nu}
- C_{\substack{eB \\ p m}}^\star \, \sigma^{\mu \nu} H^\dagger \, \ell_p \, B_{\mu \nu}, \nn
& - 2 \, C_{\substack{ee \\m prs}} \gamma_\mu e_p \, J^{e \, \mu}_{rs}-  C_{\substack{eu \\m prs}} \gamma_\mu e_p \, J^{u \, \mu}_{rs}
-  C_{\substack{ed \\m prs}} \gamma_\mu e_p \, J^{d \, \mu}_{rs}
- C_{\substack{\ell e \\ pr m s}} \gamma_\mu e_s \, J^{\ell \, \mu}_{pr}
- C_{\substack{q e \\ pr m s}} \gamma_\mu e_s \, J^{q \, \mu}_{pr}, \nn
&- C_{\substack{\ell e d q \\ p m st}}^\star \, (\ell_{p, \,j})(\overline{q_t}^j \, d_s) - C_{\substack{\ell e q u \\p m st}}^{(1),\star} \, (\ell_{p}^j) \, \epsilon_{jk} \, (\overline{u_t} \, q_s^k)
- C_{\substack{\ell e q u \\ p m st}}^{(3), \star} \, \sigma_{\mu \nu} \ell_p^j  \, \epsilon_{jk} \, (\overline{u_t} \, \sigma^{\mu \nu} q_s^k),\\
\Delta_{d,m}^{(6,B)} &= - C_{\substack{dH \\p m}}^\star \, H^\dagger \,(H^\dagger H) q_p
- C_{\substack{Hd \\m p}} \, (H^\dagger i\overleftrightarrow D_\mu H) \gamma^\mu d_p
+ C_{\substack{Hud \\m p}}^\star \, i \left[(D_\mu H)^\dagger \tilde{H} \right] \gamma^\mu u_p
- C_{\substack{dW \\p m}}^\star \, \sigma^{\mu \nu} H^\dagger \, \tau_I q_p \, W^I_{\mu \nu}, \nn
& - C_{\substack{dB \\p m}}^\star \, \sigma^{\mu \nu} H^\dagger \, q_p \, B_{\mu \nu}
- C_{\substack{dG \\p m}}^\star \, \sigma^{\mu \nu} H^\dagger \, T_A \, q_p \, G^A_{\mu \nu}
- C_{\substack{dd \\m prs}} \gamma_\mu d_p \, J^{d \, \mu}_{rs}
- C_{\substack{dd \\rs m p}} \gamma_\mu d_p \, J^{d \, \mu}_{rs}
-  C_{\substack{ed \\rs m p}} \gamma_\mu d_p \, J^{e \, \mu}_{rs},\nn
& -  C^{(1)}_{\substack{ud \\rs m p}} \gamma_\mu d_p \, J^{u \, \mu}_{rs}
- C_{\substack{\ell d \\st m p}} \, \gamma_\mu \, d_p  J^{\ell \, \mu}_{st}
-  C^{(8)}_{\substack{ud \\rs m p}} \gamma_\mu \, T_A \, d_p \, J^{u, A, \, \mu}_{rs}
- C^{(1)}_{\substack{q d \\st m p}} \, \gamma_\mu \, d_p J^{q \, \mu}_{st}
- C^{(8)}_{\substack{q d \\st m p}} \, \gamma_\mu \, T_A \, d_p \, J^{q, A, \, \mu}_{st}, \nn
& - C_{\substack{\ell e d q \\st m p}} \, q_{p, \,j} \,(\overline{\ell_s}^j \, e_t)
- C_{\substack{quqd \\st p m}}^{(1),\star} \, (\overline{u_t} \, q_s^j) \, \epsilon_{jk} \, q_p^k
- C_{\substack{quqd \\st p m}}^{(8),\star} \, (\overline{u_t} \, T^A \, q_s^j) \, \epsilon_{jk} \, T_A \, q_p^k,
\end{align}
\begin{align}
\Delta_{u,m}^{(6,B)} &= - C_{\substack{uH \\p m}}^\star \, (H^\dagger H) \, \tilde{H}^\dagger \, q_p
- C_{\substack{Hu \\m p}} \, (H^\dagger i\overleftrightarrow D_\mu H) \gamma^\mu u_p
- C_{\substack{Hud \\m p}} \, i(\tilde{H}^\dagger D_\mu \, H) \gamma^\mu d_p
- C_{\substack{uW \\p m}}^\star \, \sigma^{\mu \nu} \tilde{H}^\dagger \, \tau_I q_p \, W^I_{\mu \nu}, \nn
& - C_{\substack{uB \\p m}}^\star \, \sigma^{\mu \nu} \tilde{H}^\dagger \, q_p \, B_{\mu \nu}
- C_{\substack{uG \\p m}}^\star \, \sigma^{\mu \nu} \tilde{H}^\dagger \, T_A \, q_p \, G^A_{\mu \nu}
- C_{\substack{uu \\m prs}} \gamma_\mu u_p \, J^{u \, \mu}_{rs}
- C_{\substack{uu \\rs m p}} \gamma_\mu u_p \, J^{u \, \mu}_{rs}
-  C_{\substack{eu \\rs m p}} \gamma_\mu u_p \, J^{e \, \mu}_{rs},\nn
& -  C^{(1)}_{\substack{ud \\m p rs}} \gamma_\mu u_p \, J^{d \, \mu}_{rs}
- C_{\substack{\ell u \\st m p}} \, \gamma_\mu \, u_p \, J^{\ell \, \mu}_{st}
-  C^{(8)}_{\substack{ud \\m p rs}} \gamma_\mu \, T_A \, u_p \,  J^{d, A, \, \mu}_{rs}
- C^{(1)}_{\substack{q u \\st m p}} \, \gamma_\mu \, u_p J^{q \, \mu}_{st}
- C^{(8)}_{\substack{q u \\st m p}} \, \gamma_\mu \, T_A \, u_p  J^{q, A, \, \mu}_{st}, \nn
& - C_{\substack{quqd \\p m st}}^{(1),\star} \, (\overline{d_t} \, q_s^k) \, \epsilon_{jk} \, q_p^j
- C_{\substack{quqd \\p m st}}^{(8),\star} \, (\overline{d_t} \, T_A \, q_s^k) \, \epsilon_{jk} \, T^A \, q_p^j
- C_{\substack{\ell e q u \\stp m}}^{(1),\star} \, (q_{p}^k) \, \epsilon_{jk} \, (\overline{e_t} \, \ell_s^j)
- C_{\substack{\ell e q u \\stp m}}^{(3),\star} \, (\sigma_{\mu \nu} \, q_{p}^k) \, \epsilon_{jk} \, (\overline{e_t} \, \sigma^{\mu \nu} \ell_s^j), \\
\Delta_{\ell,m}^{(6,B),j} &= - C_{\substack{eH \\m p}} \, (H^\dagger H) e_p \, H^j
- C_{\substack{eW \\m p}} \, \sigma^{\mu \nu} \, e_p \, \tau_I \, H^j \, W^I_{\mu \nu}
- C_{\substack{eB \\m p}} \, \sigma^{\mu \nu}  \, e_p \,  H^j \,  B_{\mu \nu}
- C_{\substack{H\ell \\m p}}^{(1)} \, (H^\dagger i\overleftrightarrow D_\mu H) \gamma^\mu \ell^j_p, \nn
&- C_{\substack{H\ell \\m p}}^{(3)} \, (H^\dagger i\overleftrightarrow D^I_\mu H) \gamma^\mu \tau_I \, \ell^j_p
- C_{\substack{\ell \ell \\ m pst}} \gamma_\mu \ell^j_p \, J^{\ell \, \mu}_{st}
- C_{\substack{\ell \ell \\ st m p}} \gamma_\mu \ell^j_p \, J^{\ell \, \mu}_{st}
- C_{\substack{\ell q \\ m pst}}^{(1)} \gamma_\mu \ell^j_p \,
J^{q \, \mu}_{st}
- C_{\substack{\ell q \\ m pst}}^{(3)} \gamma_\mu \, \tau_I \, \ell^j_p \,
J^{q \, I \, \mu}_{st}, \nn
&- C_{\substack{\ell e \\ m pst}} \gamma_\mu \ell^j_p \,
J^{e \, \mu}_{st} - C_{\substack{\ell u \\ m pst}} \gamma_\mu \ell^j_p \, J^{u \, \mu}_{st}
- C_{\substack{\ell d \\ m pst}} \gamma_\mu \ell^j_p \, J^{d \, \mu}_{st}
- C_{\substack{\ell e d q \\m pst}} \, e_{p}  \, (\overline{d_s} \, q^j_t)
- C_{\substack{\ell e q u \\m pst}}^{(1)} \, e_{p} \, \epsilon^{jk} \, (\overline{q_{s,k}} \, u_t), \nn
&- C_{\substack{\ell e q u \\m pst}}^{(3)} \, e_{p} \sigma_{\mu \nu}\, \epsilon^{jk} \, (\overline{q_{s,k}} \, \sigma^{\mu \nu} u_t), \\
\Delta_{q,m}^{(6,B),j} &=
- C_{\substack{uH \\m p}} \, (H^\dagger H) u_p \, \tilde{H}^j
- C_{\substack{dH \\m p}} \, (H^\dagger H) d_p \, H^j
- C_{\substack{uW \\m p}} \, \sigma^{\mu \nu} \, u_p \, \tau_I \, \tilde{H}^j \, W^I_{\mu \nu}
- C_{\substack{dW \\m p}} \, \sigma^{\mu \nu} \, d_p \, \tau_I \, H^j \, W^I_{\mu \nu}, \nn
&- C_{\substack{uB \\m p}} \, \sigma^{\mu \nu} \, u_p \, \tilde{H}^j \, B_{\mu \nu}
- C_{\substack{dB \\m p}} \, \sigma^{\mu \nu} \, d_p \, H^j \, B_{\mu \nu}
-C_{\substack{uG \\p m}} \, \sigma^{\mu \nu} u_p \tilde{H}^j \, T_A \, G^A_{\mu \nu}
-C_{\substack{dG \\p m}} \, \sigma^{\mu \nu} d_p \, H^j \, T_A \, G^A_{\mu \nu}\nn
&- C_{\substack{H q \\m p}}^{(1)}  (H^\dagger i\overleftrightarrow D_\mu H) \gamma^\mu q^j_p
- C_{\substack{H q \\m p}}^{(3)}  (H^\dagger i\overleftrightarrow D^I_\mu H) \gamma^\mu \, \tau_I \, q^j_p
- \left(C_{\substack{qq \\ m pst}}^{(1)} + C_{\substack{qq \\ st m p}}^{(1)}\right) \gamma_\mu q^j_p  J^{q  \mu}_{st}
- \left(C_{\substack{qq \\ st m p}}^{(3)}+ C_{\substack{qq \\ m p st }}^{(3)}\right) \gamma_\mu  \tau_I  q^j_p
J^{q, I, \mu}_{st}, \nn
&- C_{\substack{\ell q \\ st m p}}^{(1)} \gamma_\mu q^j_p \,
J^{\ell, \mu}_{st}
- C_{\substack{\ell q \\ st m p}}^{(3)} \gamma_\mu  \tau_I  q^j_p
J^{\ell, I,  \mu}_{st}
- C_{\substack{q e \\ m pst}} \gamma_\mu q_p^j \,
J^{e, \mu}_{st}
- C_{\substack{q u \\ m pst}}^{(1)} \gamma_\mu q_p^j \,
J^{u, \mu}_{st}
- C_{\substack{q u \\ m pst}}^{(8)} \gamma_\mu \, T_A \, q_p^j \,
J^{u, A, \mu}_{st}, \nn
&- C_{\substack{q d \\ m pst}}^{(1)} \gamma_\mu q_p^j \,
J^{d, \mu}_{st}
 - C_{\substack{\ell e d q \\ st p m}}^\star (\overline{e_t} \ell^j_s) \, d_p
- C_{\substack{q d \\ m pst}}^{(8)} \gamma_\mu \, T_A \, q_p^j \,
J^{d, A,\mu}_{st} - C_{\substack{quqd\\ m pst}}^{(1)} \, u_p \, \epsilon^{jk} \, (\overline{q_{s,k}} d_t)
- C_{\substack{quqd\\ st m p}}^{(1)} \, (\overline{q_{s,k}} u_t) \, \epsilon^{kj} \, d_p, \nn
&- C_{\substack{quqd\\ m pst}}^{(8)} \, T^A \, u_p \, \epsilon^{jk} (\overline{q_{s,k}} \, T_A \, d_t)
- C_{\substack{quqd\\ st m p}}^{(8)} \, (\overline{q_{s,k}} \, T_A \, u_t) \epsilon^{kj} \, T^A \, d_p
- C_{\substack{\ell equ\\ st m p}}^{(1)} \, (\overline{\ell_{s,k}} e_t)  \epsilon^{kj}  u_p
- C_{\substack{\ell equ\\ st m p}}^{(3)} \, (\overline{\ell_{s,k}}  \sigma^{\mu \nu}  e_t) \epsilon^{kj} \,
 \sigma_{\mu \nu} \, u_p.
\end{align}

There are additional corrections due to the B number violating operators of $\mathcal{L}^{(6)}$.
Defining these operators as
\begin{align}
\mathcal{Q}_{\substack{duq\\ prst}} &= \epsilon^{\alpha \beta \rho} \epsilon^{jk}
(\overline{d_p^{\alpha,c}} u_r^\beta)(\overline{q_{s, \, j}^{\rho,c}} \ell_{t, \, k}), & \quad
\mathcal{Q}_{\substack{qqu\\ prst}} &= \epsilon^{\alpha \beta \rho} \epsilon^{jk}
(\overline{q_{p, j}^{\alpha,c}} q_{r, \, k}^{\beta})(\overline{u_s^{\rho,c}} e_t), \nn
\mathcal{Q}_{\substack{qqq\\ prst}} &= \epsilon^{\alpha \beta \rho} \epsilon^{jo} \epsilon^{kl}
(\overline{q_{p, j}^{\alpha,c}} q_{r, \, k}^{\beta})(\overline{q_{s, \, l}^{\rho,c}} \ell_{t, \,o}), &\quad
\mathcal{Q}_{\substack{duu\\ prst}} &= \epsilon^{\alpha \beta \rho}
(\overline{d_p^{\alpha,c}} u_r^{\beta})(\overline{u_s^{\rho,c}} e_t),
\end{align}
results in the EOM corrections of the following form
\bea
\Delta_{e,m}^{(6,\slashed{B})}  &=& -\epsilon^{\alpha \beta \rho} \left[
C_{\substack{qqu \\ prso}}^\star
\epsilon^{jk} (\overline{q_{r,k}^{\beta}} q_{p, j}^{\alpha,c}) u_s^{\rho,c}
+ C_{\substack{duu \\ prsm}}^\star
(\overline{u_r^{\beta}} d_p^{\alpha,c}) u_s^{\rho,c} \right], \nn
\Delta_{u,m \, \rho}^{(6,\slashed{B})} &=& - \epsilon^{\alpha \beta \rho} \left[
C_{\substack{duq \\ pm st}}^\star \epsilon_{jk} (\overline{\ell_t^{j}} q_s^{\beta \, k,c}) d_p^{\alpha,c}
+ C_{\substack{duu \\ p r m t}}^\star (\overline{d_p^{\alpha,c}} u_r^{\beta})^\star e_t^{c}
+ C_{\substack{qqu \\ pr m t}}^\star \epsilon^{j \, k}  (\overline{q^{\alpha,c}_{p, j}} q_{r, k}^{\beta})^\star e_t^{c}
- C_{\substack{duu\\p m st}}^\star \, d_p^{\alpha, c}(\overline{e_t} u_s^{\beta, c}) \right], \nn
\Delta_{d,m \, \rho}^{(6,\slashed{B})}  &=& - \epsilon^{\rho \beta \alpha} \left[
C_{\substack{duq \\ m rst}}^\star \epsilon^{j \, k} (\overline{q_{s,j}^{\alpha, c}} \ell_{t, \, k})^\star \, u_r^{\beta,c}
+ C_{\substack{duu \\ m rst}}^\star (\overline{u_s^{\alpha, c}} e_t)^\star \, u_r^{\beta,c} \right],
\eea
(here $\alpha,\beta,\rho$ are $\rm{SU}(3)_c$ indices) and for the $\rm SU(2)_L$ doublet fields
\bea
\Delta_{q,m, \rho}^{(6,\slashed{B}),j}&=&
-\epsilon^{\alpha \beta \rho} \left(C_{\substack{qqu\\ p m st}}^\star \epsilon^{jk} (\overline{e_t} u_s^{\beta,c}) q_{p, k}^{\alpha,c}
+C_{\substack{qqu\\ m rst}}^\star \epsilon^{jk} (\overline{u_s^{\beta,c}} e_t)^\star q_{r, k}^{\alpha,c}
+  \epsilon^j_{ \, \, k}\left[C_{\substack{duq\\ pr m t}}^\star   \ell_{t}^{k,c}
(\overline{d_p^{\alpha,c}} u_{r}^{\beta})^\star
- C_{\substack{qqq\\ pr m t}}^\star  \epsilon^{lo}  \ell_{t,o}^{c}
(\overline{q_{p,l}^{\alpha,c}} q_{r}^{\beta,k})^\star\right] \right), \nn
&-&  \epsilon^{\alpha \beta \rho} \epsilon^{jo} \epsilon^{kl}\left[C_{\substack{qqq\\ m rst}}^\star
q_{r,k}^{\alpha,c} (\overline{q_{s,l}^{\beta,c}} \ell_{t,o})^\star
- C_{\substack{qqq\\ p m st}}^\star q_{p,k}^{\alpha,c} (\overline{\ell_{t,l}} q_{s,o}^{\beta,c})\right],\\
\Delta_{\ell,m}^{(6,\slashed{B}),j} &=& - \epsilon^{\alpha \beta \rho} \left(
C_{\substack{duq\\ prs m}}^\star \, \epsilon^{jk} (\overline{u_r^\alpha} d_{p}^{\beta, c})\, q_{s,k}^{\rho,c}
+ C_{\substack{qqq\\ prs m}}^\star \, \epsilon^{jo} \, \epsilon^{kl}
(\overline{q_{r,k}^\alpha} q_{p,o}^{\beta, c}) q_{s,l}^{\rho,c} \right).
\eea
The gauge field EOM corrections are
\begin{align}
\Delta_{B,\mu}^{(6)} &= 2 \hyp_H  (H^\dagger H)
\left[ \sum_{\psi = \ell,q} \, C_{\substack{H \psi \\ pr}}^{(1)} \, J^{\psi \, \mu}_{pr}
+ \sum_{\psi = e,u,d} C_{\substack{H \psi \\ pr}} \, J^{\psi \, \mu}_{pr}
+ \frac{C_{HD}}{2} \, H^\dagger i\overleftrightarrow D_\mu H \right]
+ 2 \hyp_H  (H^\dagger \, \tau_I \, H)
\sum_{\psi = \ell,q} \, C_{\substack{H \psi \\ pr}}^{(3)} J^{\psi \, I \, \mu}_{pr}, \nn
&+\frac{4 \, C_{HB}}{g_1} \partial^\nu (H^\dagger H) B_{\nu \, \mu}
+ \frac{2 \, C_{HWB}}{g_1} [D^\nu, H^\dagger \tau H]_I W^I_{\nu \, \mu}
+ 4 \, C_{HB} \, H^\dagger H \, j_\mu + \frac{2 \, g_2}{g_1} \, C_{HWB} \, (H^\dagger \, \tau_I \, H) J^I_\mu, \nn
&+ \frac{4 \, C_{H\tilde{B}}}{g_1} \partial^\nu \left(H^\dagger H \tilde{B}_{\nu \, \mu}\right)
+ \frac{2 \, C_{H\tilde{W} B}}{g_1} [D^\nu, H^\dagger \tau H]_I \tilde{W}^I_{\nu \, \mu}
+ \frac{2 \, C_{H\tilde{W} B}}{g_1} [D^\nu, \tilde{W}_{\nu \, \mu}]_I \, H^\dagger \tau^I H, \nn
&+ \frac{2}{g_1} \left( \partial_\nu \mathcal{C}_{\substack{e  \ell B \\pr}}^{\nu \mu}
+ \partial_\nu \tilde{\mathcal{C}}_{\substack{u q B \\pr}}^{\nu \mu}
+ \partial_\nu \mathcal{C}_{\substack{d q B \\pr}}^{\nu \mu} \right),\\
\Delta_{W,\mu}^{I,(6)} &= 2 (H^\dagger \, \tau^I \, H)
\left[\sum_{\psi = \ell,q} \, C_{\substack{H \psi \\ pr}}^{(1)} \, J^{\psi \, \mu}_{pr}
+ \sum_{\psi = e,u,d} C_{\substack{H \psi \\ pr}} \, J^{\psi \, \mu}_{pr}
+ \frac{C_{HD}}{2} \, H^\dagger i\overleftrightarrow D_\mu H \right]
+ 2 \, i \, \epsilon^{IJK} \, (H^\dagger \, \tau_J \, H) \sum_{\psi = \ell,q} \, C_{\substack{H \psi \\ pr}}^{(3)} J^{\psi \, \mu}_{pr \, K}, \nn
&+ C_{\substack{H ud \\ pr}}  (\tilde{H}^\dagger \, \tau^I \, H) (\overline{u_p} \, \gamma_\mu d_r)
+ C^\star_{\substack{H ud \\ pr}}  (H^\dagger \, \tau^I \, \tilde{H}) (\overline{d_r} \, \gamma_\mu u_p)
+ \frac{4}{g_2} \, \partial^\nu (H^\dagger H) (C_{HW} W_{\nu \mu}^I + C_{H\tilde{W}} \tilde{W}_{\nu \mu}^I), \nn
& + 4 \, C_{HW} \, H^\dagger  H \, j_\mu^I
+ \frac{4 \, C_{H \tilde{W}}}{g_2} \, H^\dagger  H \, \left[D^\nu , \tilde{W}_{\nu \mu} \right]^I
+ \frac{2}{g_2} \partial^\nu \left[ (H^\dagger \tau^I H) \,  B_{\nu \mu} \, C_{HWB} +  (H^\dagger \tau^I H) \,  \tilde{B}_{\nu \mu} \, C_{H\tilde{W}B} \right], \nn
&+ \frac{2}{g_2} \, \left[D^\nu, \mathcal{C}^{\nu \mu}_{\substack{e  \ell W \\pr,I}} \right]^I
+ \frac{2}{g_2} \, \left[D^\nu, \mathcal{C}^{\nu \mu}_{\substack{d q W \\pr,I}} \right]^I
+ \frac{2}{g_2} \, \left[D^\nu, \tilde{\mathcal{C}}^{\nu \mu}_{\substack{u q W \\pr,I}} \right]^I
+2 (H^\dagger \, \tau_J \, H) \epsilon^{IJK} W_{\nu \, K} \, ( C_{HWB} B_{\nu \mu} + C_{H\tilde{W}B} \tilde{B}_{\nu \mu}), \nn
&+ \frac{6 \, C_{W}}{g_2} \, \epsilon^{IJK} (\partial^\gamma (W^{\mu \nu}_J \, W^{\nu \gamma}_K) + g_2 \epsilon_{RSK}
W_{\gamma \nu}^R \, W_{\nu \mu}^S \, W_J^\gamma)
+ \frac{2 \, C_{\tilde{W}}}{g_2} \, \epsilon^{IJK} (\partial^\gamma (\tilde{W}^{\nu \gamma}_J \, W^{\mu \nu}_K) + g_2 \epsilon_{RSJ}
\tilde{W}_{\nu \gamma}^R \, W_{\mu \nu}^S \, W_K^\gamma), \nn
&+ \frac{2\, C_{\tilde{W}}}{g_2} \, \epsilon^{IJK} (\partial^\gamma (\tilde{W}^{\mu \nu}_J \, W^{\gamma \nu}_K) + g_2 \epsilon_{RSJ}
\tilde{W}_{\mu \nu}^S \, W_{\nu \gamma}^R \, W_K^\gamma)
+ \frac{C_{\tilde{W}}}{g_2} \, \epsilon^{IJK} \epsilon_{\rho \phi}^{\, \, \, \, \, \gamma \mu} (\partial^\gamma (\tilde{W}^{\rho \nu}_J \, W^{\nu \phi}_K) + g_2 \epsilon_{RSJ}
\tilde{W}_{\nu \gamma}^S \, W_{\rho \nu}^R \, W_K^\phi), \\
\Delta_{G,\mu}^{A,(6)} &=
\frac{4 }{g_3} \, \partial^\nu (H^\dagger \, H) (C_{HG} G_{\nu \mu}^A + C_{H\tilde{G}} \tilde{G}_{\nu \mu}^A)
+ \frac{4}{g_3} \, H^\dagger  H \, \left(C_{HG} \, g_3 \, j_\mu^A  + C_{H \tilde{G}}
\left[D^\nu , \tilde{G}_{\nu \mu} \right]^A\right), \nn
& + \frac{2}{g_3} \, \left[D^\nu, \mathcal{C}^{\nu \mu}_{\substack{d q G \\pr,A}} \right]^A
+ \frac{2}{g_3} \, \left[D^\nu, \tilde{\mathcal{C}}^{\nu \mu}_{\substack{u q G \\pr,A}} \right]^A
+ \frac{6 \, C_{G}}{g_3} \, f^{ABC} (\partial^\gamma (G^{\mu \nu}_B \, G^{\nu \gamma}_C) + g_3 f_{DEC}
G_{\gamma \nu}^D \, G_{\nu \mu}^E \, A_B^\gamma), \nn
&+ \frac{2\, C_{\tilde{G}}}{g_3} \, f^{ABC} (\partial^\gamma (\tilde{G}^{\nu \gamma}_B \, G^{\mu \nu}_C) + g_3 f_{DEB}
\tilde{G}_{\nu \gamma}^D \, G_{\mu \nu}^E \, A_C^\gamma)
+ \frac{2 \, C_{\tilde{G}}}{g_3} \, f^{ABC} (\partial^\gamma (\tilde{G}^{\mu \nu}_B \, G^{\gamma \nu}_C) + g_3 f_{DEB}
\tilde{G}_{\mu \nu}^E \, G_{\nu \gamma}^D \, A_C^\gamma), \nn
&+ \frac{C_{\tilde{G}}}{g_3} \, f^{ABC} \epsilon_{\rho \phi}^{\, \, \, \, \, \gamma \mu} (\partial^\gamma (\tilde{G}^{\rho \nu}_B \, G^{\nu \phi}_C) + g_3 f_{DEB}
\tilde{G}_{\nu \gamma}^E \, G_{\rho \nu}^D \, A_C^\phi).
\end{align}
\end{widetext}

\end{document}